\magnification=1200
\baselineskip=16pt
% Plain Format macro file ----------
\catcode`@=11 \def\b@lank{ }
\newif\if@simboli\newif\if@riferimenti\newif\if@incima\newif\if@bozze
\def\bozze{\@bozzetrue \immediate\write16{stampa nome equazioni}}
\newwrite\file@simboli\def\simboli{\immediate\write16{ Genera \jobname.SMB }
\@simbolitrue\immediate\openout\file@simboli=\jobname.smb
\immediate\write\file@simboli{Simboli di \jobname}}
\newwrite\file@ausiliario\def\riferimentifuturi{
\immediate\write16{ Genera \jobname.AUX }\@riferimentitrue\openin1 \jobname.aux
\ifeof1\relax\else\closein1\relax\input\jobname.aux\fi
\immediate\openout\file@ausiliario=\jobname.aux}
\newcount\eq@num\global\eq@num=0\newcount\sect@num\global\sect@num=0
\newcount\lemm@num\global\lemm@num=0
\newif\if@ndoppia\def\numerazionedoppia{\@ndoppiatrue\gdef\la@sezionecorrente{\the\sect@num}}
\def\se@indefinito#1{\expandafter\ifx\csname#1\endcsname\relax}
\def\spo@glia#1>{} \newif\if@primasezione
\@primasezionetrue \def\s@ection#1\par{\immediate
\write16{#1}\if@primasezione\global\@primasezionefalse\else\goodbreak
\vskip\spaziosoprasez\fi\noindent
{\bf#1}\nobreak\vskip\spaziosottosez\nobreak\noindent}
\def\eqpreset#1{\global\eq@num=#1
\immediate\write16{ !!! eq-preset = #1 }     }
\def\eqlabel#1{\global\advance\eq@num by 1
\if@ndoppia\xdef\il@numero{(\la@sezionecorrente.\the\eq@num)}
\else\xdef\il@numero{(\the\eq@num)}\fi
\def\usa@getta{1}\se@indefinito{@eq@#1}\def\usa@getta{2}\fi
\expandafter\ifx\csname @eq@#1\endcsname\il@numero\def\usa@getta{2}\fi
\ifodd\usa@getta\immediate\write16
{ *** possibili riferimenti errati a \string\eqref{#1} !!!}\fi
\expandafter\xdef\csname @eq@#1\endcsname{\il@numero}
\if@ndoppia       \def\usa@getta{\expandafter\spo@glia\meaning
\la@sezionecorrente.\the\eq@num}\else\def\usa@getta{\the\eq@num}\fi
\if@simboli\immediate\write\file@simboli{  Equazione 
\usa@getta :  eqref.   #1}\fi
\if@riferimenti\immediate\write\file@ausiliario{
\string\expandafter\string\edef
\string\csname\b@lank @eq@#1\string\endcsname{\usa@getta}}\fi}
\def\eqref#1{\se@indefinito{@eq@#1}
\immediate\write16{ *****\string\eqref{#1} non definita !!!}
\if@riferimenti\relax
\else\eqlabel{#1} ***\fi \fi\csname @eq@#1\endcsname }
\def\autoeqno#1{\eqlabel{#1}\eqno\csname @eq@#1\endcsname\if@bozze
{\tt #1}\else\relax\fi}\def\autoleqno#1{\eqlabel{#1}
\leqno(\csname @eq@#1\endcsname)}
\def\titoli#1{
\@incimatrue\nopagenumbers\xdef\prima@riga{#1}
\if@incima\voffset=+30pt\headline={\if\pageno=1{\hfil}\else\hfil{\sl 
\prima@riga}\hfil\folio\fi}
\fi} \catcode`@=12

\def\frac#1#2{{#1\over {#2}}}

\def\capo{\par\noindent } 

% % end of macro file --------------

\nopagenumbers
\
\vskip 1 truecm
\capo
\centerline{\bf PHYSICALLY VALID BLACK HOLE INTERIOR MODELS}\capo
\vskip 3 truecm 
\capo
\centerline{{\bf Giulio Magli}}\capo
\vskip 1 truecm
\centerline{
Dipartimento di Matematica del Politecnico di Milano}\capo
\centerline{P.le Leonardo da Vinci 32, 20133 Milano, 
Italy}\capo
\centerline{tel: 39-2-23994538 (voice) 39-2-23994568 (fax)}\capo
\centerline{e-mail magli@mate.polimi.it}

\vskip 3 truecm\capo

{\sl New, simple models of ``black hole interiors'', namely 
spherically symmetric solutions of the Einstein field equations in 
matter matching the Schwarzschild vacuum at spacelike
hypersurfaces ``$R<2M$'' are constructed.
The models satisfy the weak energy condition 
and their matter content 
is specified by an equation of state of the elastic type.}

\vskip 1 truecm\capo
\centerline{04.70.Bw 04.20.Jb}

\vfill\eject 

Spherically symmetric, static models of stars in General Relativity 
are composed by
a regular solution of the Einstein field equations in matter
matching smoothly the Sch\-warz\-schild vacuum at a surface which
lies outside the event horizon, while
the Sch\-warz\-schild black hole is usually regarded {\it per se}  
as a singular solution of the Einstein field equations;
its physical acceptability from the astrophysical point of view
relies in the fact
that the singularity is covered by the horizon.
However, of course, the problem of singularities is far from being solved,
and it is generally believed that quantum effects should restrict the curvature
to finite values.
As a consequence, in recent years
the possibility of an alternative picture in which also black holes
have non--singular interiors has been actively investigated [1-6].
For ``interior structure'' of a (non--rotating) black hole 
one means a solution of the Einstein field equations
in matter matching smoothly the Sch\-warz\-schild solution
in a region which lies {\it inside} the Schwarzschild radius.
Of course, speculations of this kind are {\it a priori}
academic because no one can communicate to an external 
observer the discovery of a ``closed world'' inside a black hole. 
We consider, however, interesting to check whether
General Relativity allows the existence of this kind of objects
at a purely classical level.
\par
Investigating 
the possible ``interior structure'' of black holes,
one is immediately faced with
rigorous theorems [7-9] which
impose severe restrictions:
no solutions satisfying the strong
or the dominant energy condition may exist.
Therefore, the physical acceptability of such models
must be restricted to the weak energy condition only.
Models of regular black holes interiors
satisfying this condition
have been found by Mars, Mart\'{\i}n-Prats and Senovilla [6].
Recently, Borde [10] remarked that a model of this kind 
is already known from unpublished work by J. Bardeen
(see also [11]).
Borde has shown that the mechanism responsible
for avoidance of singularities
in all regular black holes is topology change.
\pageno=1
\titoli{G. Magli, Physically valid black hole interior models}
%\vfill\eject

\par
Taken together, the above mentioned results give us a quite 
complete understanding of ``regular black hole physics''.
It is, however, worth mentioning that,
if we really want to check if nature allows the existence of such objects,
we should find models satisfying an equation of state, 
so that a clear physical interpretation 
of the matter content of their source can be given.
As far as the present author is aware,
no examples of this kind are known, and our aim here
is to fill this gap.
Our starting point is that
strongly collapsed,
relativistic bodies such as neutron stars
typically exhibit elastic--solid properties 
due to a process of crystallization of matter at very high densities
(see e.g. [12]).
Therefore, there is a possible interpretation 
of isolated, anisotropic sources of strong gravitational fields
as highly collapsed, elastic solid objects.
In the present paper,
we propose a very simple class of ``elastic--solid'' 
black hole interiors.

Following Ref.[6],
we consider the spherically symmetric
line element in Ed\-ding\-ton--Finkelstein coordinates:
$$
ds^2=-e^{4\beta}\chi du^2 +2 e^{2\beta} du dr 
+r^2 ( d\theta^2 +\sin^2(\theta)d\phi^2 )\ ,\autoeqno{ed}
$$
where $\beta$ and $\chi$ are functions of $r$ only.
It is useful to represent $\chi$ as
$$
\chi (r) =1-\frac {2\mu (r)}r\ .
$$
The form \eqref{ed} of the line element is very convenient
in dealing with the black hole interior problem.
In fact, observe that
one can always use the transformation between the null coordinate $u$
and the Schwarzschild ``time'' $t$ given by
$$
dt(u,r)=du -\frac {e^{-2\beta (r)}}{\chi (r)}dr
$$
which brings the line element in its Schwarzschild form,
but this transformation is allowed only {\it separately}  
within regions in which the function $\chi$
has a constant sign.
It is worthwhile to note that, rewriting \eqref{ed} as
$$
ds^2=-e^{4\beta}\chi \left(
du -\frac {e^{-2\beta}}\chi dr\right)^2 +\frac 1\chi dr^2
+r^2 ( d\theta^2 +\sin^2(\theta)d\phi^2 )\ ,\autoeqno{ed1}
$$
the line element in Schwarzschild form is static
if $\chi$ is positive, while it is non--static 
if $\chi$ is negative.
We shall refer towards to such regions as 
to $R-$ and $T-$ regions respectively [13].
The line element \eqref{ed} can be used 
to describe spacetimes which contain both $R-$ and $T-$ regions.
\par
To proceed further,
we assume as a source 
an isotropic elastic medium.
The energy--momentum tensor of an elastic--solid sphere 
has been thoroughly discussed in Ref.[14].
In particular, it was shown that 
a spherically symmetric, relativistic 
body can be characterized by three
quantities $\epsilon, p, \Omega$.
In terms of such quantities the eigenvalues
of the energy--momentum tensor may be written as
$$
\eqalign{
\lambda_0 &=-\epsilon \ ,\cr
\lambda_1 &=p+2\Omega \ ,\cr
\lambda_2 &=\lambda_3 = p-\Omega \ .
}\autoeqno{atv}
$$
In $R-$regions, $\lambda_0$ corresponds to the time--like eigenvector
and therefore $\epsilon$ is the energy density,
while $\lambda_1$ is the radial stress, 
which the sum of two terms:
the isotropic part $p$ and the quantity $\Omega$ which
is a measure of the response 
of the body to strains that change its shape without changing
is volume: if $\Omega$ vanishes, the elastic medium is simply a perfect fluid.
The eigenvalue $\lambda_2$ always corresponds to a space--like eigenvector,
and therefore is the tangential stress, while,
in $T-$regions, 
the role of $\lambda_0$ and $\lambda_1$ is reversed.
\par
The Einstein field equations may be written as follows:
$$
\eqalign{
\mu^\prime &=4\pi r^2 \epsilon \ ,\cr
\chi \beta^\prime 
&=2\pi r(\epsilon + p+2\Omega)\ ,
\cr 
2\chi \left(
p^\prime + 2\Omega^\prime +6\frac \Omega{r}\right)
&=-(\epsilon +p +2\Omega)(4\chi \beta^\prime +\chi^\prime )
\ ,
}
\autoeqno{1}
$$ 
where a dash denotes the derivative with respect to $r$.
This system is composed by three equations for the 
five unknowns ($\chi ,\beta ,\epsilon ,p, \Omega$)
and therefore it is not closed
until we supplement it with an equation of state.
Generally speaking, to specify the
equation of state for an isotropic elastic body
one gives the energy density as a function of the three 
invariants of the strain tensor.
The stresses may then be calculated as 
derivatives of the energy with respect to such invariants
(in the case of a perfect fluid, this 
obviously amounts to say that the equation of state 
relates energy and specific volume $v$, and that the pressure
may be obtained as 
the derivative of the energy with respect to $v$).
In the particular case of spherical symmetry,
it has been shown [14] that assigning the state 
equation of the body consists
in giving the energy density
as a function of only two invariants of the strain tensor,
one of them being the specific volume (so that
the derivative of $\epsilon$ with respect to it governs the
isotropic response $p$) and the other one being the 
quadratic invariant of the deformation,
the ``response'' associated to this invariant being $\Omega$.
\par
The weak energy condition requires
$$
T_{\mu\nu}v^\mu v^\nu \geq 0 \ ,\autoeqno{wecc}
$$
for any timelike vector $v^\mu$.
Taking into account \eqref{atv},
the above inequality leads to
$$
\epsilon \geq 0\ ,\ \epsilon+p+2\Omega \geq 0\ ,\ \epsilon +p -\Omega \geq 0\ ,
$$
in $R-$regions, and to
$$
p+2\Omega \leq 0\ ,\ \epsilon +p+2\Omega \leq 0\ ,\ \Omega \leq 0\ ,
$$ 
in $T-$regions.
\par
A very simple way to simplify
the above conditions is to consider
that particular class of elastic materials satisfying 
$$
\epsilon+p+2\Omega =0\ .\autoeqno{consti}
$$
In fact, the weak energy condition for such materials
reduces to 
$$
\epsilon \geq 0\ ,\ \Omega \leq 0\ ,\autoeqno{wec1}
$$
in {\it any} region.
We stress that the above choice, besides being very convenient from the
mathematical point of view, has a clear physical meaning.
It is, in fact, equivalent to a ``constitutive'' partial
differential equation
for $\epsilon$. The solution of such equation 
identifies a well--defined class of elastic materials
which depends on the choice of an arbitrary function of one variable only
(we refer the reader to Ref. [14] for details).
This arbitrariness may be used to assign the ``on shell''
value $\epsilon =\epsilon (r)$.
The system \eqref{1} then becomes closed and reduces to
$$
\eqalign{
\mu^\prime &=4\pi r^2 \epsilon \ ,\cr
\beta^\prime &=0\ ,\cr
\epsilon^\prime &=
6\frac \Omega{r}\ .
}\autoeqno{1w}
$$
Therefore, the function $\chi$ is given by
$$
\chi (r):= 1-\frac{8\pi}r\int_0^r s^2\epsilon(s) ds\ ,
$$ 
the function $\beta$ is a constant which 
may be set equal to zero without loss of generality,
and the line element assumes the simple form 
$$
ds^2=-\chi (r) du^2 +2 du dr 
+r^2 ( d\theta^2 +\sin^2(\theta)d\phi^2 )\ .\autoeqno{sol}
$$
This solution may be seen as 
a ``variation of the mass'' of the 
Schwarzschild solution and has
many interesting features.
For example, it is a Kerr--Schild geometry and it arises naturally
as the spherically symmetric limit of a recently found class of sources of 
the Kerr metric [15].
It arises also in completely different contexts,
for example in general--relativistic electron models (see e.g. 
[16] and references therein) or in vacuum polarization [5].
\par
Let us now consider the conditions which have to be imposed 
on the free function $\epsilon$ in order to obtain a regular 
black hole model.\par
First of all, the weak energy condition \eqref{wec1}
requires $\epsilon$
to be a non--negative function, while $\Omega =r\epsilon^\prime /6$ 
must be negative, so that $\epsilon$ must be a decreasing
function of $r$.
The regularity of the spacetime at $r=0$
requires $\mu \sim r^3$ as $r$ tends to zero,
and therefore $\epsilon$ must be finite there.
Finally, in order to obtain a black--hole model, 
we want to match smoothly the interior solution \eqref{sol}
with the Schwarzschild solution at a {\it space--like}
hypersurface $r=R$.
The matching between two metrics 
is smooth (no surface distributions of matter--energy
arise) if the first and the second fundamental form
are continuous at the matching surface.
Continuity of the metric requires
the mass of the vacuum solution to be given by
$$
M:=\mu(R)\ .
$$ 
To obtain a black--hole model,
we require $R$ to belong to a $T-$region:
$$
1-\frac {2M}R<0 \ .\autoeqno{bh1}
$$
Continuity of the second fundamental form requires
vanishing of the energy density since $r=R$ is a space--like hypersurface.
The energy density is 
$-(p+2\Omega)$ in a $T-$region,
but finally in our model this equals 
$\epsilon$, so that
we require $\epsilon (R)=0$.
\par
Regarding the structure of the solutions, it is easy to check
that the function $\chi$
must vanish at at least one value $r_o$ of $r$,
because $\chi(r)$ is continuous and satisfies $\chi(0)=1$, $\chi(R)<0$.
This is a characteristic shared by all regular black holes interiors
satisfying the weak energy condition [6].  
A nice feature of our solutions is that
there exist always only one value $r_o$,
as a graphic comparison between $r$ and the monotone increasing function 
$2\mu (r)$ immediately shows.
A simple calculation also shows
that the Riemann invariants are finite, and therefore
the solution is non--singular and $r_o$ is a Cauchy horizon.
At the horizon, the Pleba\'{n}ski type of the energy--momentum tensor
degenerates from $[2S_1 - S_2 - T]$
to $[2S - 2N]$ 
(the timelike and the radial spacelike eigenvector degenerate into 
a double null eigenvector).
It may, however, be easily shown that the weak energy condition 
remains satisfied there.
\par
The maximal analytic extension of our solutions is similar
to that already well known for others ``black hole interiors''
(see e.g. [5-6]).
It closely resembles that of the Reissner--Nordstr\"om
spacetime with the key difference that the singularity at $r=0$
is replaced by the matter--filled region.

It may be instructive to discuss some 
explicit examples.
Considering a power-law distribution of density 
$$
\epsilon =\alpha \left[1-\left(\frac rR \right)^n\right]\ ,\autoeqno{dens}
$$
(where $\alpha$ and $n$ are positive quantities),
the Schwarzschild mass is
$$
M=\frac {4n}{3(n+3)} \pi R^3 \alpha \ ,
$$
and the solution is a black hole interior provided that
$$
R> R_{\rm min}:=\left[\frac {3(n+3)}{8\pi n\alpha}\right]^{1/2}\ ,
$$
therefore, at fixed $n$, the minimal required value of $r$ 
at the matching hypersurface decreases as $\alpha$
increases.
The solution is given by \eqref{sol} with
$$
\chi (r)=1-
\frac{2M}{nR}\left(\frac rR\right)^2 \left[n+3-3\left(\frac rR\right)^n
\right]\ .
$$
\par
Since the central value $\alpha=\epsilon (0)$ is really the 
density at the centre because $r=0$ always belongs
to a $R-$region, it seems reasonable to assume 
it around the typical values for neutron stars,
namely of order $10^{15}$ g/cm$^3$.
Considering, for instance,
the case of a linear function $\epsilon$ ($n=1$ 
in \eqref{dens})
we can calculate $R_{\rm min}$.
Since 
we are using relativistic units,
we can express it in Kilometers
(it is about $25$ Km), although it would be
more correct to express it in time--units (about $8.2\times 10^{-5}$ seconds)
because it is in fact an instant of time, belonging to a $T-$region.
In any case, of course, the unique quantity which is
really measurable by an external observer 
is the mass of the Schwarzschild black--hole.
Assuming $R$ to be $26$ Km one obtains this mass
to be $M\approx 8.5 M_{\odot}$, with a
Schwarzschild radius
$2M\approx 28$ Km.
\capo

\vskip .3 truecm
\par\noindent

The author thanks Jerzy Kijowski for useful discussions and comments.

\vfill\eject

\item{[1]} 
E. Poisson and W. Israel,
``Structure of the black hole nucleus.''
Class. Quantum Grav. {\bf 5} (1988) L201--L205.

\capo
\item{[2]} 
E. Poisson and W. Israel,
``Internal structure of black holes.'' 
Phys. Rev. {\bf D 41}, (1990), 1796--1809. 

\capo
\item{[3]} 
V. P. Frolov, M.A. Markov and V. F. Muchanov,
``Through a black hole into a new universe?''
Phys. Lett. {\bf B 216} (1989) 272--276.

\capo
\item{[4]} 
V. P. Frolov, M.A. Markov and V. F. Muchanov,
``Black holes as possible sources of closed and semiclosed worlds.''
Phys. Rev. {\bf D 41} (1990) 383--394.

\capo
\item{[5]}
I. Dymnikova,
``Vacuum non singular black hole.''
Gen. Rel. Grav. {\bf 24} (1992) 235--242.

\capo
\item{[6]} 
M. Mars, M.M. Mart\'{\i}n-Prats and J.M.M. Senovilla,
``Models 
of regular Schwarzschild black holes satisfying weak energy conditions.'' 
Class. Quantum Grav. {\bf 13} (1996) L51--L58.

\capo
\item{[7]}
S. Hawking and G.F.R. Ellis, 
{\it The large--scale structure of space--time.} 
Camb. Un. Press, Cambridge 1973.

\item{[8]}
T.W. Baumgarte and A.D. Rendall,
``Regularity 
of spherically symmetric static solutions of the Einstein equations.'' 
Class. Quantum Grav. {\bf 10} (1993) 327--332.

\item{[9]} M. Mars, M.M. Mart\'{\i}n-Prats and J.M.M. Senovilla,
``The $2m\leq r$ property of spherically symmetric
static space--times.''
Phys. Lett. {\bf A218} (1996) 147--150. 

\item{[10]}
A. Borde,
``Regular black holes and topology change.''
Phys. Rev. {\bf D 55} (1997) 7615--7617.

\item{[11]}
A. Borde,
``Open and closed universes, initial singularities, and inflation.''
Phys. Rev. {\bf D 50} (1994) 3692--3702.

\item{[12]} 
S.L. Shapiro and S.A. Teukolsky,  
{\it Black Holes, White Dwarfs and Neutron Stars.}
Wiley, New York (1983). 
\capo

\item{[13]} Y.D. Zeldovich and I.D. Novikov, 
{\it Relativistic Astrophysics I.} 
Chicago Un. Press, Chicago (1971).
\capo

\item{[14]} G. Magli and J. Kijowski,  
``A generalization of the relativistic equilibrium equations for a 
non--rotating star.'' 
Gen. Rel. Grav. {\bf 24} (1992) 139--158.
\capo

\item{[15]} G. Magli,  
``Kerr-Schild gravitational fields 
in matter and the Kerr-interior problem 
in general relativity.''
J. Math. Phys. {\bf 36} (1995) 5877--5896.
\capo

\item{[16]} L. Herrera and V. Varela,
``Finite 
self-energy of pointlike sources for the Reissner-Nordstr\"om metric.'' 
Gen. Rel. Grav. {\bf 28} (1996) 663--678.\capo

\end